\documentclass[prl,twocolumn,draft,amsmath,showpacs]{revtex4}
\usepackage{graphics}
\usepackage{graphicx}

\begin{document}

\bibliographystyle{prsty}
\input epsf

\title{Unconventional Charge Ordering in Na$_{0.70}$CoO$_{2}$ below 300 K}

\author{J. L. Gavilano, D. Rau, B. Pedrini, J. Hinderer, H. R. Ott, S.
M. Kazakov and J. Karpinski}

\affiliation{ Laboratorium f\"{u}r Festk\"{o}rperphysik,
ETH-H\"{o}nggerberg, CH-8093~Z\"{u}rich, Switzerland }


\begin{abstract}

We present the results of measurements of the dc-magnetic
susceptibility $\chi(T)$ and the $^{23}$Na-NMR response of
Na$_{0.70}$CoO$_{2}$ at temperatures between 50 and 340 K. The
$\chi(T)$ data suggest that for $T > 75$ K, the Co ions adopt an
effective configuration of Co$^{3.4+}$.  The $^{23}$Na-NMR response
reveals pronounced anomalies near 250 and 295 K, but no evidence for
magnetic phase transitions is found in $\chi(T)$.
Our data suggest the onset of a dramatic change in the Co 3d-electron
spin dynamics at 295 K. This process is completed at 230 K. Our
results maybe interpreted as evidence for either a tendency to
electron localization or an unconventional charge-density wave
phenomenon within the Co 3d electron system near room temperature.

\end{abstract}
\maketitle


The discovery of high-$T_{c}$ superconducting cuprates has initiated
an enhanced interest in transition-metal oxides with layered
structures, among them layered alkali-metal cobalt oxides.  In analogy
with the high-$T_{c}$ cuprates it was argued that cobalt-oxide layers
may result in unusual physical properties of these compounds,
including unconventional superconductivity\cite{Tanaka1994}.  Indeed,
recent experiments\cite{Takada2003} suggest that the hydration of
nominal Na$_{0.70}$CoO$_{2}$ induces superconductivity below a
critical temperature $T_{c} \approx 5$ K. Other recent investigations
confirm that the physical properties of Na$_{x}$CoO$_{2}$ are quite
puzzling\cite{Fujimoto2003,Kobayashi2003,Waki2003,Wang2003,Wang2003a,Motohashi2003,Sugiyama2003,Singh2003}.

Na$_{x}$CoO$_{2}$, with $x < 1$, crystallizes in a hexagonal structure
with $P6_{3}/mmc$ space group symmetry\cite{Balsys1996,jorgensen2003},
a structure consisting of Na layers, intercalated between sheets
formed by triangular arrangements of CoO$_{6}$ octahedra.  Tunnels
with an effective radius of 0.81 \AA $ $ connect the Na positions and
allow the Na ions (ionic radius 0.97 \AA) to move, providing some
degree of ionic conductance above room temperature\cite{Delmas1981}.

We report on measurements of the dc-magnetic susceptibility and
$^{23}$Na-NMR on polycrystalline samples of Na$_{0.70}$CoO$_{2}$ at
temperatures between 50 and 300 K and in external magnetic fields of
2.815 and 7.049 T. Above 295 K the Co ions are intermediate valent
with an average configuration Co$^{3.4+}$.  Drastic changes in the Co
3d electron system occur below 295 K. This may be interpreted as
either an enhanced tendency to electron localization with decreasing $T$ or the
formation of an unconventional charge-density-wave state.

Our sample consisted of randomly oriented powder of nominal
Na$_{0.70}$CoO$_{2}$, which was obtained by heating a stoichiometric
mixture of Co$_{3}$O$_{4}$ (Aldrich, 99.995$\%$) and Na$_{2}$Co$_{3}$
(Aldrich, 99.995$\%$) in air for 15 h at 850 $^{\textrm{o}}$C. To
ensure the oxygen content the powder was annealed for 2 h at an
oxygen pressure of 500 bar at 500 $^{\textrm{o}}$C. X-ray powder
diffraction showed that the resulting material was of single phase
with lattice parameters $a$ = 2.826(1) and $c$ = 10.897(4) \AA.
Neutron powder diffraction experiments performed on samples prepared
in a similar way yielded a composition of Na$_{x}$CoO$_{2}$ with x =
0.7 $\pm$ 0.04\cite{Kazakov2003}.
The dc-magnetic susceptibility was measured using the moving-sample
technique with a commercial SQUID magnetometer and for the NMR
measurements we used standard spin-echo techniques.

\begin{figure}[t]
 \begin{center}
  \leavevmode
  \epsfxsize=0.8\columnwidth \epsfbox {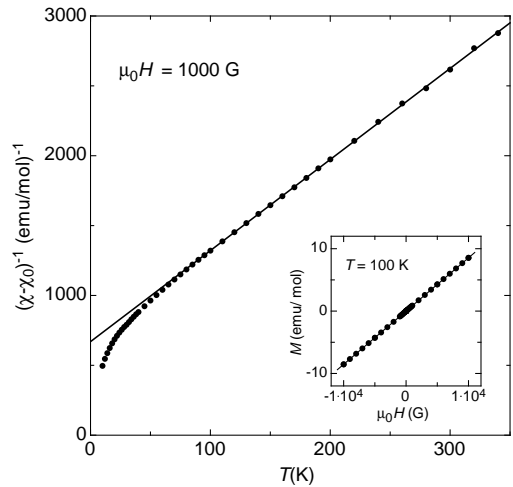}

\caption{ The inverse  magnetic susceptibility as a function
of temperature for Na$_{0.7}$CoO$_{2}$. The solid line represents
a Curie-Weiss type fit. Inset: Magnetization as a function of the
external magnetic field at 100 K.} \protect 
\label{Figure1}
\end{center}
\end{figure}

In Fig.  1 we display the temperature dependence of the inverse
magnetic susceptibility $(\chi - \chi_{0})^{-1}$, whereby $\chi_{0} =
1.25 \cdot 10^{-4}$ emu/mol, obtained by fitting the
experimental data by a curve which includes a Curie-Weiss law and a
temperature-independent contribution ($\chi_{0}$), at temperatures
above 100 K.
Between 75 and 340 K, the Curie-Weiss type behavior is emphasized by
the solid line.  The inset of Fig.  1 shows the magnetization $M$ as a
function of the applied field $H$ at 100 K. It reflects the
paramagnetic response of the magnetic moments that are responsible for
the behavior of $\chi(T)$.  The solid straight line in Fig.  1 is
compatible with an effective paramagnetic moment $p_{eff}$ = 1.1
$\mu_{B}$ per Co ion, and a paramagnetic Curie temperature
$\theta_{\mathrm{p}}$ of -103 K,
in fair agreement with previous results for Na$_{0.75}$CoO$_{2}$ (see
Ref.  8).
The value of $p_{eff}$ can be interpreted as evidence for i) all the
Co ions are in an intermediate valent Co$^{3.4+}$ state, or ii) their
configurations are Co$^{4+}$ (S = 1/2) and Co$^{3+}$ (S = 0) with an 
approximate ratio of 2:3, 
if it is assumed that $g = 2$ and the orbital moments are quenched. 
Considering the Na content, this ratio is reasonable.

\begin{figure}[t]
\begin{center}
\leavevmode
\epsfxsize=0.8\columnwidth \epsfbox {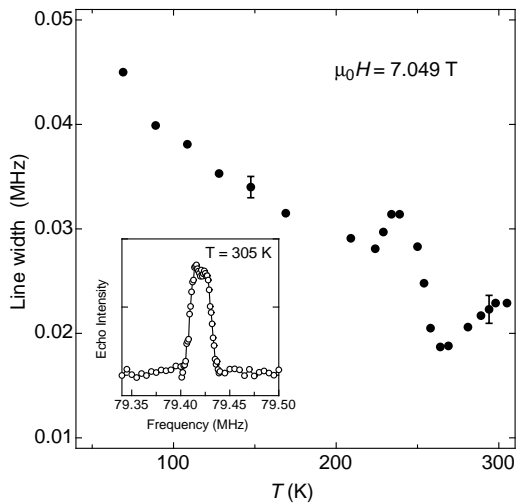}

\caption{ Main frame: Temperature dependence of the linewidth (FWHM)
of the $^{23}$Na-NMR central transition $-1/2 \leftrightarrow 1/2$ for
Na$_{0.7}$CoO$_{2}$.  Inset: The central transition of the
$^{23}$Na-NMR spectrum at $T = $ 305 K and $\mu_{0} H = $ 7.049 T. }
\protect
\label{Figure2}
\end{center}
\end{figure}

The $^{23}$Na-NMR spectra were measured in fixed external magnetic
fields of 2.815 and 7.049 T. In the inset of Fig.  2 we display an
example of a signal measured in a field of 7.049 T and at $T = $ 305
K, which is attributed to the $^{23}$Na nuclear-Zeeman central
transition $-1/2 \leftrightarrow 1/2$.  The shape of this line (see
also Fig.  4) is as expected for a randomly oriented powder, broadened
by second-order quadrupolar effects\cite{Carter1977}.  This is
corroborated by the fact that at high temperatures the NMR linewidth
scales linearly with the inverse of the applied magnetic field. 
Analyzing the $^{23}$Na NMR lineshape of the central line at
temperatures above 295 K yields a quadrupolar frequency $\nu_{Q} =
e^{2}qQ/(2h)$ of 1.65 $ \pm $ 0.2 MHz.  Here $eq$ is the largest
component of the electric field gradient EFG at the Na nuclear site
and $eQ$ is the quadrupolar moment of the Na nucleus.  Note that above
room temperature the EFG's are the same at all Na sites.

In the main frame of Fig.  2 we display the variation of the
$^{23}$Na-NMR linewidth (FWHM) with temperature.  Below 200 K it
reflects, analogous to $\chi(T)$, the increase of the magnetization
due to the Co$^{4+}$ ions with decreasing temperature.  A prominent
anomaly is observed in FWHM($T$) near 250 K. Since we find no
corresponding anomaly in $\chi(T)$, this feature cannot involve
magnetic degrees of freedom and, as we argue below, it most
probably is related to structural aspects.


\begin{figure}[t]
\begin{center}
\leavevmode
\epsfxsize=0.8\columnwidth \epsfbox {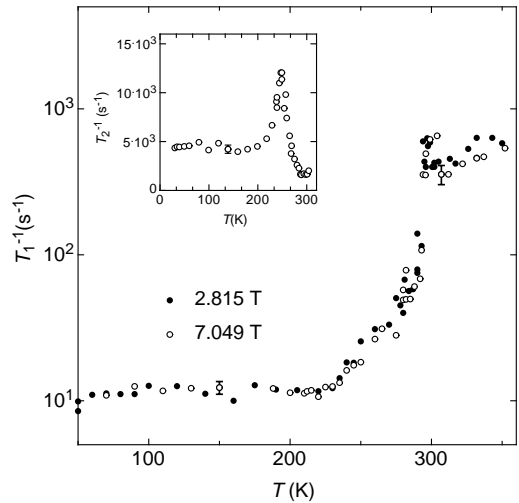}

\caption{ Main frame: Spin-lattice relaxation rate as a function of
$T$ measured at two different external magnetic fields.
Inset: Spin-spin relaxation rate as a function of $T$, measured
in an external magnetic field of 2.815 T.} \protect
\label{Figure3}
\end{center}
\end{figure}


In Fig.  3 we present the temperature dependence of the spin-lattice
relaxation rate $T_{1}^{-1}(T)$, measured in fields of 2.815 and 7.049
T. The values of $T_{1}^{-1}$ were extracted from standard
fits\cite{Simmons62} to the nuclear magnetization recovery curves of
the $^{23}$Na central nuclear Zeeman transition, after the application
of a long comb of $rf$ pulses.  Below 230 K, $T_{1}^{-1}$ is
approximately $T-$independent, as in the case of spin-lattice relaxation
driven by flips of paramagnetic moments\cite{Narath73}. 
With increasing temperature we observe, near 230 K, the onset of an
increase of the relaxation rate $T_1^{-1}$.  This onset is accompanied
by a significant anomaly in the $T$-dependence of the spin-spin
relaxation rate $T_2^{-1}$, which is shown in the inset of Fig.  3. 
Together, these two relaxation-rate features provide evidence for a
crossover phenomenon or a phase transition.  Close to 250 K, the value
of $T_{2}^{-1}$ is of the order of the NMR linewidth (Fig.  2) which,
therefore, is dominated by $T_{2}$ effects.

With increasing $T$, the slope $\textrm{d}T_1^{-1}/\textrm{d}T$
increases and an almost discontinuous enhancement of $T_1^{-1}$, by a
factor of 20, occurs at 295 K. In view of the smooth evolution of
$\chi(T)$, this rapid change in $T_{1}^{-1}(T)$ manifests a phase
transition which obviously does not involve magnetic degrees of
freedom.  No anomaly in the spin-spin relaxation rate is observed at
295 K. The order of magnitude difference of $T_{1}^{-1}$ below and
above 295 K seems too large to be caused by a structural phase
transition.  In an effort to interpret the observed relaxation
features, we focus on possible changes within the Na and Co subsystems,
respectively.


\begin{figure}[t]
\begin{center}
\leavevmode
\epsfxsize=1\columnwidth \epsfbox {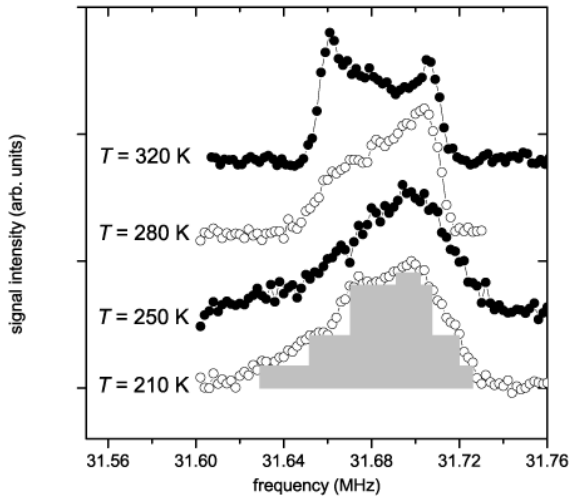}

\caption{
$^{23}$Na-NMR central transitions of Na$_{0.7}$CoO$_{2}$
 measured in an external field of 2.815 T at different temperatures.
 The gray boxes represent rough sketches of the different contributions to the
 NMR line at 210 K. The line at 320 K has a single contribution.
} \protect
\label{Figure4}
\end{center}
\end{figure}

First we consider the $T-$induced changes of the quadrupolar features
of the $^{23}$Na-NMR line (see Fig.  4).  The lineshape exhibits clear
$T-$induced changes, but no evidence for ``motional
narrowing'', which is expected if an order-disorder transition
occurred at 295 K, such as, $e.g.$, in Ba$_{2}$In$_{2}$O$_{5}$ at
1075 $^{\textrm{o}}$C\cite{Adler1994}.  Thus, a Na order-disorder
transition seems unlikely to be the reason for the observed changes of
the NMR spectrum at and below 295 K.
Recent experiments on Na$_{0.7}$CoO$_{2}$ searched for, but did not
find, temperature-induced changes in the occupation of the 2 Na sites
below room temperature\cite{Gavilano2003}.  Likewise, no appreciable
changes in the occupation of the 2 Na sites below 300 K for
Na$_{0.61}$CoO$_{2}$ were reported in ref. 12.  Hence the
transition at 295 K is most likely due to a $T-$induced variation in
the Co subsystem.

Considering the results for $\chi(T)$, the homogeneity of the EFG and
the very high and temperature independent spin-lattice relaxation rate
above room temperature, we conclude that the Co ions are in an
intermediate valent state of 3.4+ above 295 K. The electron
transitions from the Co 3d orbitals to conduction band states is,
compared to the NMR frequencies, very fast.  The rapid decrease of the
spin-lattice relaxation rate observed below 295 K and the
corresponding changes in the NMR spectrum may be interpreted as
evidence for a partial electron localization, $i.e.$, the
characteristic frequency $\tau_{c}^{-1}$ (with $\tau_{c}$ the
correlation time) of the valence fluctuations decreases to values well
below the NMR Larmor frequencies as the temperature is reduced.  At
lower temperatures a ``quasi'' mixed-valent phase with Co$^{4+}$ (S =
1/2), and Co$^{3+}$ (S = 0) ions in a ratio of 2 to 3 is adopted.  The
peak in $T_{2}^{-1}(T)$ seems to indicate that $\tau_{c}$ reaches a
value of the order of 10$^{-4}$ s, $i.e.$, of the order of $T_{2}$,
near 250 K\cite{deSoto96}.  A complete and static charge localization
cannot occur in Na$_{0.7}$CoO$_{2}$ because this would imply an
insulating phase, contrary to the experimental
observations\cite{Batlogg2003}.  This partial localization should be
distinguished from case of a ``slowing down'' of the spin flips of
localized moments, which is unlikely to occur.  If below 200 K,
$\tau_{c}^{-1}$ is well below the NMR frequencies (``slow-motion
regime'') and if we assume that the spin-lattice relaxation is driven by
the independent flips of ``slow'' and localized spins then, neglecting
spin diffusion\cite{Benoit63},
\begin{equation}
    T_{1}^{-1} =
A \cdot \tau_{c}/(1+\omega^{2}\tau_{c}^{2}) \approx 
A/(\omega^{2}\tau_{c}) ,
    \label{eq:Eq1}
\end{equation}
where $A$ is a constant.  Since, as seen in Fig.  3, $T_{1}^{-1}(T)$
is field and frequency independent it would follow that $\tau_{c}
\propto H^{-2}$, whereas for these high temperatures ($\mu_{B}H \ll
k_{B}T$) a field-independent $\tau_{c}$ is expected.  More likely is
the scenario described above, where the ionic moments form and decay
by the motion of electrons, and these changes drive the nuclear
relaxation.


In trying to interpret the features at 250 K, we first focus on the
NMR spectrum at 210 K, whose lineshape is not drastically different
from the one at 250 K, but exhibits some distinct shoulders.  Although
these features are not very prominent, they are perfectly reproducible
and their relative positions scale with $1/H$ (data not shown) as
expected for quadrupolar features of the $^{23}$Na-NMR central line. 
We identify four contributions, represented by the gray rectangles in
Fig.  4, which we attribute to four inequivalent Na sites in the
Co-ion environment.  The quadrupolar frequencies of the four Na sites
are of the order of 2.6, 2, 1.5 and less than 0.7 Mhz, respectively.
These subtle changes of the NMR spectrum may be due to either a slight
re-arrangement of the static or, on the time scale of the NMR
experiments, very slowly varying positions of the Na ions. 
Alternatively, very slow Co charge fluctuations cannot be excluded. 
Temporally varying Na environments are mainly determined by the actual
configuration of nearest neighbor Co$^{4+}$ ions.  At low
temperatures, this configuration is stable over a sufficiently
long time period, such that they appear as static in the NMR
measurements and provide the shoulders in the line at 210 K. At higher
temperatures, however, this feature may be lost.

Since the changes of the $^{23}$Na-NMR spectrum between 210 and 250 K,
discussed above, are rather subtle, we performed 2D exchange
$^{23}$Na-NMR experiments\cite{Ernst1987,Dolinsek1998}.  The results
of these experiments will be discussed in detail
elsewhere\cite{Gavilano2003}, but we mention here that our experiments
detected (data not shown) small changes of the Larmor frequencies of
Na nuclei, in the temperature range between 240 - 260 K and on a time
scale $\tau_{mix}$, typically smaller than 10$^{-4}$ s, but
$\tau_{mix}^{-1} \ll \nu_{L}$, the NMR Larmor frequency.  At first
sight, this observation maybe interpreted as evidence for a restricted
and slow motion of Na ions near 240 K, which decreases again at higher
temperatures.  This seems rather counter-intuitive and therefore
probably indicates that the frequency changes are due to changes in
the Na environments, $i.e.$, in the Co-oxide layer.  Our 2D NMR
experiments cannot distinguish between genuine Na ionic motion and
temporal changes of the Na environments.

Another scenario consistent with our experimental results is the
formation of an unconventional charge density wave in the Co subsystem
at 295 K. The system remains metallic at low
temperatures\cite{Maki2002}, but part of the Fermi surface is lost. 
In the density-wave state, the Co 3d-electron spin dynamics may change
and account for the rapid decrease of the spin-lattice relaxation,
that we observe below 295 K. The charge density wave may also lead to a
slight rearrangement of the Na ions around 250 K.

Support for our claims are obtained from the results of a structural
investigation\cite{Gavilano2003} of a single crystalline sample of
Na$_{0.7}$CoO$_{2}$ by $X-$ray diffraction.  The comparison of
diffraction patterns taken at 220 and 300 K show the presence of an
additional structural feature at 220 K. Recent density functional
calculations suggested\cite{Singh2000} that charge as well as spin
ordering is nearly degenerate with the paramagnetic state in
Na$_{0.5}$CoO$_{2}$.

In conclusion, from the results of our dc-susceptibility and NMR
measurements, we infer that below 295 K the Co $3d-$electron system
within the CoO layers is affected by at least one phase transition. 
Two scenarios seem possible: (a) a partial charge ordering phenomenon
involving Co$^{3+}$ and Co$^{4+}$ upon decreasing temperature,
completed only near 230 K; and (b) an unconventional charge density
wave within the Co subsystem develops.  The system remains metallic at
all temperatures.

This work was financially supported
by the Schweizerische Nationalfonds zur F\"{o}rderung der
Wissenschaftlichen Forschung.




\end{document}